\documentstyle[12pt]{article}

\newcommand{\be}{\begin{equation}}
\newcommand{\ee}{\end{equation}}
\newcommand{\bea}{\begin{eqnarray}}
\newcommand{\eea}{\end{eqnarray}}

\newcommand{\tA}{\tilde A}
\newcommand{\tB}{\tilde B}
\newcommand{\tC}{\tilde C}
\newcommand{\tD}{\tilde D}

\newcommand{\ta}{\tilde a}
\newcommand{\tb}{\tilde b}
\newcommand{\tc}{\tilde c}
\newcommand{\td}{\tilde d}

\newcommand{\nn}{\nonumber}

\newcommand{\RR}{\rangle}
\newcommand{\LL}{\langle}

\newcommand{\RL}{\RR\LL}

\makeatletter
\newdimen\normalarrayskip              
\newdimen\minarrayskip                 
\normalarrayskip\baselineskip
\minarrayskip\jot
\newif\ifold             \oldtrue            

\makeatother
\newlength{\extraspace}
\newlength{\extraspaces}
\input{tcilatex}

\begin{document}

\addtolength{\baselineskip}{3mm}

\thispagestyle{empty}

\begin{flushright}
\baselineskip=12pt
quant-ph/0001105\\
\hfill{  }\\ 
\end{flushright}
\vspace{.5cm}

\begin{center}
\baselineskip=24pt

{\Large {\bf {Quantum Anti-Cloning}}}\\[15mm]

\baselineskip=12pt

{\bf David D. Song\footnote{{\bf E-mail: {\tt d.song@qubit.org}}} and 
Lucien Hardy\footnote{{\bf E-mail: {\tt l.hardy@qubit.org}}}} \\[%
8mm]
{\it Centre for Quantum Computation\\[0pt]
Clarendon Laboratory, University of Oxford\\[0pt]
Parks Road, Oxford OX1 3PU, U.K.} \vspace{5cm}

{\sc Abstract}

\begin{minipage}{15cm}
\baselineskip-12pt

We derive the transformation for the optimal universal quantum anti-cloner which
 produces two anti-parallel outputs for 
a single input state.  The fidelity is shown to be 2/3 which is same as the measurement fidelity.
We consider a probabilistic quantum anti-cloner and show quantum states can be anti-cloned
 exactly with non-zero probability and its efficiency is higher than the efficiency of distinguishing 
between the two states.

\end{minipage}
\end{center}

\vfill

\newpage

\pagestyle{plain} \setcounter{page}{1}

\section{Introduction}
Possessing superposition and entanglement properties, quantum information has revealed many interesting features 
which classical information has no counterpart.
Unlike classical information, quantum information cannot be duplicated, i.e. unknown quantum state cannot be copied 
exactly \cite{zurek}.
However a universal quantum (approximate) cloner has been introduced \cite{buzek,ekert} which takes an unknown quantum state and generates multiple copies, 
with fidelity 5/6 in case of two copies as outputs regardless of an input.  
 Another feature of quantum information has been discovered  recently \cite{gisin1}
 where it was shown that more quantum information can be gained from two anti-parallel spins than from two 
 parallel ones, i.e. one can measure the spin direction $|{\bf n}\RR$ with better fidelity when two qubits are
 in anti-parallel, $|{\bf n},-{\bf n}\RR$, than in parallel, $|{\bf n},{\bf n}\RR$. 

In this paper, we consider a universal quantum anti-cloner which takes an unknown quantum state just
 as in quantum cloner but 
its output as one with the same copy while the second one with opposite spin direction to the input state.
For the Bloch vector, an input  ${\bf n}$, quantum anti-cloner would have the input as  
$\frac{1}{2}({\bf 1}+{\bf n}\cdot {\bf \sigma})$,
then it generates two outputs, $\frac{1}{2}({\bf 1}+\eta {\bf n}\cdot {\bf \sigma})$ 
and $\frac{1}{2}({\bf 1}-\eta{\bf n}\cdot{\bf \sigma})$,   where $0\leq \eta \leq 1$ 
is the shrinking factor 
and the fidelity is defined as $F = \LL {\bf n} |\rho^{({\rm out})}|{\bf n}\RR = \frac{1}{2} (1+\eta )$.
 If spin flipping were allowed then anti-cloner would have the same fidelity as the regular cloner since 
one could clone first then flip the spin of the second copy.
However spin flipping of an unknown state is not allowed in quantum mechanics.
Consider a spin-flipping of an unknown state,   
\be
\left( \begin{array}{c}
e^{\frac{-i\varphi}{2}} \cos \frac{\theta}{2}  \\
e^{\frac{i\varphi}{2}} \sin \frac{\theta}{2} 
\end{array}  \right)
\rightarrow
\left( \begin{array}{c}
-e^{\frac{-i\varphi}{2}} \sin \frac{\theta}{2}  \\
e^{\frac{i\varphi}{2}} \cos \frac{\theta}{2} 
\end{array}  \right)
\ee
This transformation can be done only by an anti-unitary operation where 
the anti-unitary transformation, $V$, satisfies the following two conditions 
\bea
(i) & &  |\LL \psi|\phi\RR | = |\LL \psi^{\prime}|\phi^{\prime}\RR| \nonumber \\
(ii) & & V \left( a|0\RR + b|1\RR \right) = a^* V|0\RR + b^* V|1\RR
\label{antiuni}\eea
where $V|\psi\RR \rightarrow |\psi^{\prime}\RR$ and $V|\phi\RR \rightarrow |\phi^{\prime}\RR$.

In sect. 2, we derive a unitary transformation for a optimal universal quantum anti-cloner
 where we obtain 2/3 for fidelity.  This value is equal to the fidelity of measurement which is for a given single unknown state, 
how precisely one can determine its state \cite{popescu}.
In sect. 3, we show that the quantum state can be anti-cloned exactly with non-zero probability. 
In case of two states, the probability of exact anti-cloning is higher than the probability 
of distinguishing between the two states. 
 We conclude with discussions on further prospects on related issues.

\section{Universal quantum anti-cloning}
In this section, we study the unitary transformation with optimal fidelity for a universal quantum anti-cloner. 
 Let us consider an input state 
$|{\bf n}\RR = \alpha |0\RR + \beta |1\RR$ 
such that for an input density matrix,
\be
\rho^{({\rm in})}= \left( \begin{array}{cc}
|\alpha|^2 & \alpha\beta^* \\
\beta\alpha^* & |\beta|^2 
\end{array} \right)
=\left( \begin{array}{cc}
1+n_z & n_x-in_y \\
n_x + in_y & 1-n_z 
\end{array} \right)
\ee
the output density matrix yields the first particle same as the input while the second one with 
opposite spin direction as follows
\be
\rho^{({\rm out})}_1 = \frac{{\bf 1} + \eta {\bf n}\cdot {\bf \sigma}}{2} = \frac{1}{2} \left( \begin{array}{cc}
1+\eta n_z & \eta(n_x - in_y) \\
\eta (n_x + in_y) & 1- \eta n_z \end{array} \right) 
\label{rho1}\ee
\be
\rho_2^{({\rm out})} = \frac{{\bf 1} - \eta {\bf n}\cdot {\bf \sigma}}{2} = \frac{1}{2} \left( \begin{array}{cc}
1-\eta n_z & -\eta(n_x - in_y) \\
-\eta (n_x + in_y) & 1+ \eta n_z \end{array} \right) 
\label{rho2}\ee
We want to consider the constraints in order to satisfy the output density matrices (\ref{rho1},\ref{rho2}) with 
maximum fidelity, i.e. $\eta$.
The conditions (\ref{rho1}) and (\ref{rho2}) imply that the two output density matrices as symmetric 
except its spin direction which are opposite to each other.  
We also impose the universality constraint that the fidelity deos not depend on the input state $|{\bf n}\RR$.

Let us consider the following general transformation,
\bea
|0\RR|Q_{23}\RR &\rightarrow& a|00\RR|A\RR + b|01\RR|B\RR + c|10\RR|C\RR + d|11\RR|D\RR \nn \\
|1\RR|Q_{23}\RR &\rightarrow& \ta |11\RR|\tA\RR +\tb |10\RR|\tB\RR + \tc |01\RR|\tC\RR  + \td |00\RR|\tD\RR
\label{transf}\eea
where $|Q_{23}\RR$ is the state to be anti-cloned and the initial ancilla state and the ancillas, $|A\RR, \cdots, |\tD\RR$,  are normalised but not necessarily orthogonal.
After following the transformation (\ref{transf}) for the input state $|{\bf n}\RR$, we have the following reduced density 
matrices after tracing out 23 and 13, respectively, 
\bea
\rho_1 &=& \{\left( |a|^2 + |b|^2\right) |\alpha|^2  + \left( a\td^* \LL \tD|A\RR + b\tc^* \LL \tC|B\RR\right) \alpha\beta^* \nn \\
&+& \left( \tc b^* \LL B |\tC\RR + \td a^* \LL A |\tD\RR\right) \beta\alpha^* + \left( |\tc|^2 + |\td|^2\right) |\beta|^2 \;\;\;\} |0\RL 0| \nn \\
&+& \{ \left(ac^* \LL C|A\RR + bd^* \LL D|B\RR \right) |\alpha|^2 + \left( a\tb^* \LL \tB|A\RR + b\ta^* \LL \tA|B\RR\right) \alpha\beta^* \nn \\
&+& \left( \tc d^* \LL  D|\tC\RR + \td c^* \LL C|\tD\RR \right) \beta\alpha^*  +\left( \tc\ta^*\LL \tA |\tC\RR + \td\tb^* \LL \tB |\tD\RR\right) |\beta|^2\;\} | 0\RL 1| \nn  \\
&+&\{ \left( ca^* \LL A|C\RR + db^* \LL B|D\RR\right) |\alpha|^2 \nn +\left( c\td^* \LL \tD|C\RR + d\tc^* \LL \tC|D\RR \right) \alpha\beta^* \nn \\
&+& \left( \ta b^* \LL B|\tA\RR + \tb a^* \LL A |\tB\RR\right) \beta\alpha^* +\left( \ta\tc^* \LL \tC |\tA\RR + \tb\td^* \LL \tD |\tB\RR \right) \beta^2\} |1\RL 0| \nn \\ 
&+&\{ \left( |c|^2 + |d|^2\right) \alpha^2 +\left( c\tb^* \LL \tB|C \RR + d\ta^* \LL \tA|D\RR\right) \alpha\beta^* \nn \\
&+&\left( \ta d^* \LL D|\tA\RR + \tb c^* \LL C |\tB\RR\right) \beta\alpha^* +\left( |\ta|^2 + |\tb|^2\right) |\beta|^2 \;\} |1\RL 1|
\label{rh1}\eea
and
\bea
\rho_2 &=& \{ \;\;\; \left(|a|^2 + |c|^2 \right) |\alpha|^2 +\left( a\td^* \LL \tD|A\RR + c\tb^* \LL \tB|C\RR\right) \alpha\beta^* \nn \\
&+&\left( \tb c^* \LL C |\tB\RR + \td a^* \LL A|\tD\RR\right) \beta\alpha^* +\left( |\tb|^2 + |\td|^2\right) |\beta|^2 \;\;\} |0\RL 0| \nn \\
&+& \{\left( ab^* \LL B|A\RR + cd^* \LL D|C\RR\right) |\alpha|^2 + \left(a\tc^*\LL \tC|A\RR + c\ta^* \LL \tA|C\RR \right) \alpha\beta^* \nn \\
&+&\left( \tb d^* \LL D|\tB\RR + \td b^* \LL B|\tD\RR\right) \beta\alpha^* +\left(\tb\ta^* \LL \tA |\tB\RR + \td\tc^* \LL \tC |\tD\RR\right) |\beta|^2 \;\} |0\RL 1\RR \nn \\
&+& \{\left( ba^* \LL A|B\RR + dc^* \LL C|D\RR \right) |\alpha|^2 + \left( b\td^* \LL \tD|B\RR + d\tb^* \LL \tB|D\RR\right) \alpha\beta^* \nn \\
&+&\left( \ta c^* \LL C|\tA\RR + \tc a^* \LL A|\tC\RR \right) \beta\alpha^* +\left( \ta\tb^* \LL \tB |\tA\RR + \tc\td^* \LL \tD |\tC\RR\right)|\beta|^2\;\} |1\RL 0| \nn \\
&+& \{\left( |b|^2 + |d|^2\right) |\alpha|^2 + \left( b\tc^* \LL \tC|B\RR + d\ta^* \LL \tA|D\RR \right) \alpha\beta^* \nn \\
&+&\left( \ta d^* \LL D|\tA\RR + \tc b^* \LL B|\tC\RR\right) \beta\alpha^* +\left(|\ta|^2 + |\tc|^2\right) |\beta|^2\;\;\} |1\RL 1|
\label{rh2}\eea
We want to consider constraints for $\rho_1$ and $\rho_2$ in (\ref{rh1},\ref{rh2}) to be same as $\rho_1^{({\rm out})}$ and $\rho_2^{({out})}$ 
in (\ref{rho1},\ref{rho2}) with maximum value for $\eta$.
Let us write the coefficients as follows
\be
a=|a|e^{i\delta_a} \; , \; b =|b|e^{i\delta_b} \; , \; c=|c|e^{i\delta_c} \; , \; d=|d|e^{i\delta_d}
\ee
and likewise for tilded cases.  Also we could write $\LL A|B\RR=|\LL A|B\RR|e^{i\delta_{AB}}$ and others are similarly defined. 
First, there are normalisation conditions to be satisfied  for the transformation  (\ref{transf}),   
\bea
|a|^2 + |b|^2 + |c|^2 + |d|^2 &=& 1 \nn \\
|\ta|^2 + |\tb|^2 + |\tc|^2 + |\td|^2 &=& 1 
\label{normal}\eea
and the orthogonality
\be
a^* \td \LL A|\tD \RR + c^* \tb \LL C| \tB \RR + b^* \tc \LL B| \tC \RR + d^* \ta \LL D| \tA \RR = 0 
\label{orth}\ee
Comparing $n_z$ terms in $\rho_1^{({\rm out})}$ and $\rho_2^{({\rm out})}$, we get the following constraints from (\ref{rh1},\ref{rh2})  
\be
|a| = |d| \;\; , \;\;\; |\ta| = |\td| 
\label{adad}\ee
\be
a\td^* \LL \tD|A\RR + b\tc^* \LL \tC|B\RR - c\tb^* \LL \tB|C\RR - d\ta^* \LL \tA|D\RR =0
\label{nz}\ee
and
\be
\eta =  |b|^2 - |c|^2 = 2|b|^2 + 2|a|^2 -1
\label{eta1}\ee
where the last relation in (\ref{eta1}) results from  (\ref{normal}) and (\ref{adad}) 
Next, Comparing $n_x$ and $n_y$ terms yields
\bea
\eta     &=& {\rm Re}[a^* \tb \LL A|\tB\RR + b^* \ta \LL B|\tA\RR ]   \label{eta2}\\
        &=& {\rm Re}[\tc a^* \LL A|\tC\RR + \ta c^* \LL C|\tA\RR]  \label{eta3}
\eea
and also the following must be satisfied. 
\bea
 &  &{\rm Im}[a^* \tb \LL A|\tB\RR + b^* \ta \LL B|\tA\RR ] =0 \label{con1}\\
&  &{\rm Im} [\tc a^* \LL A|\tC\RR + \ta c^* \LL C |\tA\RR] =0  \\
  & & b\td^*\LL \tD|B\RR + d\tb^* \LL \tB|D\RR =0  \\
  &  &ca^* \LL A|C\RR + db^* \LL B|D\RR =0  \\
  &  &\ta \tc^* \LL \tC |\tA\RR + \tb \td^* \LL \tD |\tB\RR =0 \\
   & &c\td^* \LL \tD|C\RR + d\tc^* \LL \tC|D\RR =0  \\
  &  & a^* b \LL A|B\RR + c^* d\LL C|D\RR =0  \\
  &  & \tb^* \ta \LL \tB |\tA\RR + \td^* \tc \LL \tD |\tC\RR =0 \label{con2}
\eea
For the transformation (\ref{transf}), we could also impose the constraint such that the output reduced 
density matrices do not change under $|0\RR \leftrightarrow |1\RR$, then the following is true,
\be
|a| = |\ta| \;\; , \;\; |b|=|\tb | \;\; , \;\; |c| = |\tc |
\ee
From (\ref{eta1},\ref{eta2},\ref{eta3})
\bea
\eta &=& |a||b|{\rm Re}[e^{i(\delta_a - \delta_{\tb} + \delta_{A\tB })}|\LL A|\tB\RR | + e^{i(\delta_b-\delta_{\ta} + \delta_{B\tA})} |\LL B|\tA\RR |] \label{plus} \\
   &=& -|a||c|{\rm Re}[e^{i(\delta_{\tc}-\delta_a + \delta_{A\tC })} |\LL A|\tC\RR | + e^{i(\delta_{\ta}-\delta_c + \delta_{C\tA})} |\LL C|\tA\RR |]  \label{minus}
\eea
then the maximum $\eta$ can be obtained when Re part in (\ref{minus}) is maximum, i.e. 2.  Therefore with (\ref{eta1}), following conditions can be obtained,
\be
|a|^2+|c|^2 = \frac{1-\eta}{2}  \;\; , \;\; |a||c|=\frac{\eta}{2}
\ee
then  
\bea
 (|a|-|c|)^2 &=& |a|^2+|c|^2 -2|a||c| \nonumber \\
              &=& \frac{1-\eta}{2} - \eta \geq 0  \nn\\
   & \Rightarrow & \eta \leq \frac{1}{3}
\eea
Therefore the maximum of $\eta$ is 1/3.
For $\eta = \frac{1}{3}$, 
\bea 
|\ta |= |a| &=& \sqrt{\frac{1}{6}} \;\;\;\;  , \;\;\;\; |\tb | = |b| = \sqrt{\frac{1}{2}}  \\
|\tc |= |c| &=& \sqrt{\frac{1}{6}} \;\;\;\; , \;\;\;\; |\td |=  |d| = \sqrt{\frac{1}{6}}
\eea
and the minus sign for (\ref{minus}) and (\ref{plus}) can be satisfied with the following phase choice. 
\be
\delta_{c}=\delta_{\tc} = \pi  \;\; , \;\; \delta_b = \delta_{\tb} = \cos^{-1}(\frac{1}{\sqrt 3})
\ee
while all other phases, $ \delta_{a},\cdots,\delta_{A\tB},\cdots$, vanish.
The ancillas satisfying the constraint (\ref{orth}),(\ref{nz}),(\ref{con1}-\ref{con2}) can be of the following form,   
\bea
|A\RR &=& (1,0,0,0) \;\; , \;\; |\tA\RR = (0,1,0,0) \\
|B\RR &=& (0,1,0,0) \;\; , \;\; |\tB\RR = (1,0,0,0) \\
|C\RR &=& (0,1,0,0) \;\; , \;\; |\tC\RR = (1,0,0,0) \\
|D\RR &=& (0,0,1,0) \;\; ,\; \; |\tD\RR = (0,0,0,1)
\eea
with the usual basis $|00\RR, |01\RR, |10\RR, |11\RR$.
Then the fidelity $\frac{2}{3}$ can be obtained with the following transformation
\bea
|0\RR |Q\RR &\rightarrow & \sqrt{\frac{1}{6}} |0000\RR + \sqrt{\frac{1}{2}}{\rm exp}[i \cos^{-1}(\frac{1}{\sqrt 3})]|0101\RR- \sqrt{\frac{1}{6}}|1001\RR + \sqrt{\frac{1}{6}}|1110\RR \\
|1\RR |Q\RR &\rightarrow & \sqrt{\frac{1}{6}}|1101\RR + \sqrt{\frac{1}{2}}{\rm exp}[i \cos^{-1}(\frac{1}{\sqrt 3})] |1000\RR - \sqrt{\frac{1}{6}}|0100\RR + \sqrt{\frac{1}{6}}|0011\RR
\eea
Note that the fidelity of universal anti-cloner, $F_{OACM}$, is same as the measurement fidelity, which is 2/3.
Hence one way to implement optimal anti-cloning is to 
 measure the unknown input state and prepare two qubits with opposite spin directions.
It is also implied that fidelity of spin flipping also should be bounded from below by 2/3,
 i.e. $F_{OSFM} \geq F_{SFM}^{\prime} = \frac{2}{3}$,
 since after anti-cloning, one can throw away the first qubit 
and will be left with the second qubit which has opposite direction to the input state.  
Therefore, the following holds,
\be
F_{OACM} \leq F_{OSFM}
\label{rel1}\ee
In \cite{gisin1,hillery}\footnote{ In \cite{hillery}, the term {\it Universal NOT-gate} was used rather than spin flipping},
 it was claimed that the optimal spin flipping is achieved with 2/3 fidelity including classical measurement.
Due to this classical information, one can prepare additional qubit as the original input state (i.e. opposite to the output)
 which implies 
$F_{OACM} \geq F_{ACM}^{\prime} = \frac{2}{3}$, therefore 
\be
F_{OACM} \geq F_{OSFM}
\label{rel2}\ee
Therefore, equality between $F_{OACM}$ and $F_{OSFM}$ holds.

\section{Probabilistic quantum anti-cloning}
There is another type of imperfect cloning, a probabilistic cloner.  Duan and Guo \cite{guo} showed that 
there can be a unitary transformation such that linearly independent states can be cloned perfectily, with non-zero 
probability.
Can anti-cloning be done probabilistically, i.e. 
can we find a unitary transformation such that $|{\bf m}_i\RR|0\RR \stackrel{{\rm Prob} \neq 0}{\longrightarrow} |{\bf m}_i\RR|-{\bf m}_i\RR\; , \; i=1,\cdots, n$, 
can be achieved.  
In order to show it, we follow Duan and Guo's method \cite{guo} with the following transformation,
\be
U(|{\bf m}_i\RR|0\RR |Y_0\RR) = \sqrt{f} |{\bf m}_i\RR|-{\bf m}_i\RR|Y_0\RR + \sum_{j=1}^n a_{ij}|Q_{12}^{(j)}\RR|Y_j\RR
\label{UU}\ee
where $|Y_0\RR$ and $|Y_j\RR$ are orthonormal  probe, 
such that whether cloning was successful or failed can be known, and $|Q_{12}^{(j)}\RR$ are normalised.
Taking inner product of (\ref{UU}), we get
\be
\left[ \LL {\bf m}_i | {\bf m}_j\RR \right] = f \left[  \LL {\bf m}_i|{\bf m}_j\RR\LL -{\bf m}_i|-{\bf m}_j\RR \right] + [a_{ij}][a_{ji}^*] 
\label{AEBECC}\ee
where we take $[\cdot ]$ to be a matrix. 
For any $n$-vector ${\bf k}=(k_1,\cdots,k_n)$,
we can write ${\bf k} [\LL {\bf m}_i|{\bf m}_j\RR ] {\bf k}^{\dagger} = \LL K | K \RR$ where $|K\RR \equiv k_1 |{\bf m}_1\RR + \cdots + k_n |{\bf m}_n\RR$.
Since $|K\RR$ is a quantum state (linear combination of $|{\bf m}_i\RR$s), its norm is always greater than or equal to zero.
It is zero only when $|K\RR$ itself is zero.  
If $|{\bf m}_1\RR,\cdots,|{\bf m}_n\RR$ are linearly independent, 
then $|K\RR$ is never zero for any $n$-vector $(k_1,\cdots,k_n)$.
Therefore when $|{\bf m}_i\RR$ are linearly independent,  $[\LL {\bf m}_i|{\bf m}_j\RR ]$ is positive definite.
Due to continuity, $[\langle {\bf m}_i | {\bf m}_j\rangle] - 
f[ \langle {\bf m}_i|{\bf m}_j\rangle\langle -{\bf m}_i|-{\bf m}_j\rangle ]  $ is  also positive
definite with sufficiently small $f$. Therefore $[\langle {\bf m}_i
| {\bf m}_j\rangle - f [\langle {\bf m}_i|{\bf m}_j
\rangle\langle -{\bf m}_i|-{\bf m}_j\rangle ] $
can be diagonalised and $[a_{ij}][a_{ji}^*]$ can be chosen such that  (\ref{AEBECC})
is satisfied. Therefore there exists a unitary operator $U$ such that (\ref
{UU}) is satisfied.

Consider the following general unitary transformation,
\bea
U \left( |{\bf m}_1\RR \right) &=& \sqrt{f} |{\bf m}_1\RR |-{\bf m}_1\RR |Y_0\RR + \sqrt{1-f}|Q\RR|Y_1\RR \nonumber \\
U \left( |{\bf m}_2\RR \right) &=& \sqrt{f} |{\bf m}_2\RR |-{\bf m}_2\RR |Y_0\RR + \sqrt{1-f}|Q\RR|Y_1\RR 
\eea
where $|Y_0\RR$ and $|Y_1\RR$ are orthonormal and $|Q\RR$ are normalised.  Then a
 cloning efficiency $f$ for probabilistic quantum anti-cloner can be obtained as follows,
\be
f \leq \frac{1-|\LL {\bf m}_1|{\bf m}_2\RR|}{1-|\LL{\bf m}_1|{\bf m}_2\RR|
 |\LL -{\bf m}_1 | -{\bf m}_2\RR|} = \frac{1-|\LL {\bf m}_1|{\bf m}_2\RR|}
{1-|\LL {\bf m}_1|{\bf m}_2\RR|^2}
\label{fidelity}\ee
The equality in (\ref{fidelity}) holds if $\LL {\bf m}_1 |{\bf m}_2\RR$ and $\LL -{\bf m}_1 |-{\bf m}_2\RR$ are real and 
positive which can be achieved by redefining these states by multiplying them by a phase.
Therefore, the probabilistic anti-cloner has the same efficiency as in the Duan and Guo's regular cloner.
One can see that the above probabilistic quantum anti-cloner can be generalised
 to clone $\mu=(L,M)$ copies for $L$ regular copies and 
$M$ copies of opposite spin direction.
For $n=2$ case, as $\mu \rightarrow \infty$, the bound (\ref{fidelity}) 
approaches the probability of distinguishability given by $1-|\LL {\bf m}_1 | {\bf m}_2 \RR |$
for given two states $|{\bf m}_1\RR$ and $|{\bf m}_2\RR$  \cite{dieks}.
 In \cite{hardy}, it was shown that the no-signalling condition restricts the number of states 
that can be cloned in a given Hilbert space.  Following the same argument, one can show that if PQACM can 
clone $N+1$ or more states in a $N$-dimensional Hilbert space,then faster-than-light signalling can be achieved.  Therefore
 no-signalling condition imposes a constraint such that probabilistic quantum anti-cloner cannot clone more than $N$ states.

Let us  consider the following simple example.
We  take $|{\bf m}_1\RR = |0\RR$, $ |{\bf m}_2\RR = \cos \theta |0\RR + \sin\theta |1\RR$,
$|Y_0\RR = |0\RR$, $|Y_1\RR = |1\RR$ , $|Q_{12}\RR = |00\RR $
then with maximum efficiency (\ref{fidelity}) of 
\be
f = \frac{1-\cos\theta}{1-\cos^2 \theta}
\ee
we can find the unitary operator $U = \sum_{i=1}^8 |N_i\RR \LL M_i | $
where $|M_1\RR = |000\RR,|M_2\RR=|001\RR$, $ \cdots$ ,$|M_8\RR =  |111\RR$ and $|N_i\RR$'s are as follows
\be
|N_1\RR = \frac{1}{\sqrt{1+\cos \theta}} |010\RR + \frac{\sqrt{\cos \theta}}{\sqrt{1+\cos \theta}}  |001\RR
\label{B1}\ee
\bea
|N_2\RR &=& -\frac{\cos \theta}{\sqrt{1+\cos \theta}}|000\RR 
+ \frac{\cos \theta (\cos \theta -1)}{\sin\theta \sqrt{1+\cos\theta}} |010\RR  
-\frac{\sin\theta}{\sqrt{1+\cos\theta}} |100\RR   \nonumber \\
&  &+ \frac{\cos\theta}{\sqrt{1+\cos\theta}} |110\RR 
+ \frac{\sqrt{\cos\theta}(1-\cos\theta)}{\sin\theta\sqrt{1+\cos\theta}} |001\RR
\label{B2}\eea
and $|N_3\RR,\cdots,|N_8\RR$ are chosen as orthonormal states to (\ref{B1}) and (\ref{B2}).
One can see $UU^{\dagger}=U^{\dagger}U={\bf 1}$ and can easily check that $U$ yields $|0\RR|1\RR$
 with the maximum efficiency given in (\ref{fidelity}).

With a similar argument, one can show the spin flipping,
 $|{\bf m}\RR \rightarrow |-{\bf m}\RR$, can be done probabilistically, i.e.
\bea
U|{\bf m}_1\RR |B_0\RR &=& \sqrt{F} |-{\bf m}_1\RR |B_1\RR + \sqrt{1-\xi_1}|Q\RR \nonumber \\
U|{\bf m}_2\RR |B_0\RR &=& \sqrt{F} |-{\bf m}_2\RR |B_2\RR + \sqrt{1-\xi_2} |Q\RR
\label{anti}\eea
can be shown to exist.
When $|B_1\RR$ and $|B_2\RR$ are orthogonal, one can identify $|-{\bf m}_1\RR$ 
and $|-{\bf m}_2\RR$ and can 
prepare as many states as one wants and its efficiency bound is
same as distinguishability between the two states.

\section{Discussions}
We have considered two types of quantum cloning for two anti-parallel outputs.
In probabilistic cloning, for two input states, the
anti-cloning efficiency is higher than the efficiency of distinguishing between the two states.
On the other hand, in case of  deterministic cloning,
 the fidelity of universal anti-cloner and the fidelity of measurement are above equal to 2/3.
 In other words, one could measure the input state and prepare two anti-parallel qubits (or as many as one wants) 
and this would have the same fidelity as in universal anti-cloner.  The question of why universal anti-cloner has 
the same fidelity as the fidelity of measurement does not seem to have an immediate explanation.

In \cite{cerf}, it was shown that, unlike as in the classical case, quantum conditional entropy, which is the information
 about B which cannot be gained by measuring A, can have negative values.  
This negativity of entropy has been puzzling and its exact physical meaning has been questioned.
In an anology with particle physics, it has been suggested \cite{cerf} that 
anti-qubits may be useful in describing quantum information processes where
anti-qubits were introduced as qubits
traveling backward in time \cite{bennett}.

\end{document}